\documentclass[pra,twoside,showpacs,twocolumn]{revtex4}
\usepackage{times}
\usepackage{amssymb}
\usepackage{amsmath}
\usepackage{graphicx}
\usepackage{epsfig}
\usepackage{dcolumn}
\usepackage{color}
\usepackage{bm}

\begin{document}

\title{Interaction of a quantum well with squeezed light: Quantum statistical properties}
\author{Eyob A. Sete and H. Eleuch }
\affiliation{Institute for Quantum Science and Engineering  and
Department of Physics and Astronomy, Texas A$\&$M University,
College Station, TX 77843-4242}
\date{\today}

\begin{abstract}
We investigate the quantum statistical properties of the light
emitted by a quantum well interacting with squeezed light from a
degenerate subthreshold optical parametric oscillator. We obtain
analytical solutions for the pertinent quantum Langevin equations in
the strong coupling and low excitation regimes. Using these
solutions we calculate the intensity spectrum, autocorrelation
function, quadrature squeezing for the fluorescent light. We show
that the fluorescent light exhibits bunching and quadrature
squeezing. We also show that the squeezed light leads to narrowing
of the width of the spectrum of the fluorescent light.
\end{abstract}
\maketitle
\section{Introduction}
Interaction of electromagnetic radiation with atoms has led to
interesting quantum features such as antibunching and squeezing. In
particular, interaction of two-level atoms with squeezed light has
extensively been studied by many authors
\cite{Gardiner86,Erenso02,Alebachew06}. These studies show that the
squeezed light modifies the width of the spectrum of the incoherent
light emitted by the atom. On the other hand, cavity QED in
semiconductor systems has been the subject of interest in connection
with its potential application in optoelectronic devices
\cite{shiedls07,Baas04,Eleuch10,Eleuch08,Eleuch04,Giacobino02}. For
example, such optical systems hold potential in realization of
optical devices that exhibits exceptional properties such as
monomode luminescence with high gain allowing the realization of
thresholdless laser. The quantum properties of the light emitted by
a quantum well embedded in a microcavity has been studied by several
authors \cite{Karr04,Qatrtropani05,Eleuch08a}. Unlike antibunching
observed in atomic cavity QED, the fluorescent light emitted by the
quantum well exhibits bunching \cite{Erenso03,Vyas00}. In the strong
coupling regime--when the coupling frequency between the exciton and
photon is larger than the relaxation frequencies of the medium and
the cavity--the intensity spectrum of the exciton-cavity system has
two well-resolved peaks representing two plaritons resonance
\cite{Chen95,Wang97}. In the experimental setting, Weisbuch
 \textit{et al}. \cite{Weisbuch92} demonstrated exciton-photon mode splitting
in a semiconductor microcavity when the quantum well and the optical
cavity are in resonance. Subsequent experiments on exciton-photon
coupling confirmed normal mode splitting and oscillatory emission
from exciton microcavities \cite{Pau95,Jacobson95}.

\begin{figure}[t]
\includegraphics{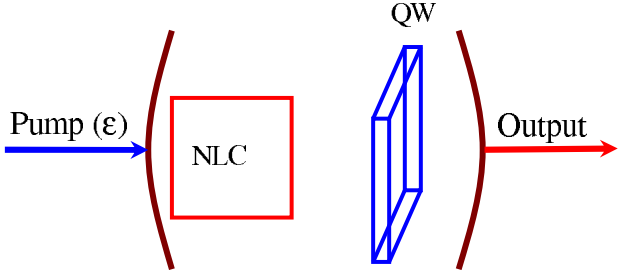}
\caption{Schematic representation of a driven cavity containing a
nonlinear crystal (NLC) and a quantum well (QW).} \label{fig1}
\end{figure}

In this work, we study the effect of the squeezed light generated by
a subthreshold degenerate parametric oscillator (OPO) on the
squeezing and statistical properties of the fluorescent light
emitted by a quantum well in a cavity. The system is outlined in
Fig. \ref{fig1}. Degenerate OPO operating below threshold is a
well-known source of squeezed light \cite{Lugiato83,Milburn83}. We
explore the interaction between this light and a quantum well with a
single exciton mode placed in the OPO cavity. Our analysis is
restricted to the weak excitation regime where the density of
excitons is small so that the interactions between an exciton and
its neighbors can be neglected. Further, to gain insight into the
physics we investigate the dynamics of the fluorescent light emitted
by the quantum well in the strong coupling regime, which amounts to
keeping the leading terms in the photon-exciton coupling constant
$g$. We show that the fluorescent light exhibits bunching and
quadrature squeezing. The former is due to the fact that two or more
excitons in the quantum well can be excited by absorbing cavity
photons. This implies there is a finite probability that two photons
can be emitted simultaneously. We also show that the squeezed light
leads to narrowing of the width of the spectrum of the fluorescent
light.

We obtain the solution of the quantum Langevin equation for a cavity
coupled to vacuum reservoir. The resulting solution, in the strong
coupling limit, is used to calculate the intensity, spectrum, second
order correlation function and quadrature squeezing of the
fluorescent light.

\section{Hamiltonian and equations of evolution}
We consider a system composed of a semiconductor quantum well and a
degenerate parametric oscillator operating below threshold. In a
degenerate parametric oscillator, a pump photon of frequency
$2\omega_{0}$ is downconverted into a pair of identical sinal
photons of frequency $\omega_{0}$. The signal photons are highly
correlated and this correlation is responsible to the reduction of
noise below the vacuum level. Such a system produces a maximum
intracavity squeezing of 50$\%$. In a quantum well, the
electromagnetic field can excite an electron from the filled valance
band to the conduction band thereby creating a hole in the valance
band. The electron-hole system possesses bound states which is also
called exciton states analogous to the hydrogenic states or more
precisely to the positronium bound states. We assume that the
density of the excitons is small so that exciton-exciton interaction
is negligible. The Hamiltonian describing the parametric process and
interaction between exciton and cavity mode in the rotating wave
approximation and at resonance is given by
\begin{equation}
H=\frac{i\varepsilon }{2}
(a^{\dagger2}-a^{2})+ig(a^{\dagger}b-ab^{\dagger})+H_{\text{loss}}.
\end{equation}
Here $a$ and $b$, considered as boson operators, are the
annihilation operators for the cavity and exciton modes,
respectively; $g$ is the exciton cavity mode coupling;
$H_{\text{loss}}$ is the Hamiltonian associated with the dissipation
of the cavity and exciton modes by vacuum reservoir modes.
We assume here that the amplitude of the field $\varepsilon $ that
drives the cavity is real and constant. The quantum Langevin
equations of the system taking into account the cavity dissipation
$\kappa $ and the exciton spontaneous emission $\gamma$ can be
written as
\begin{equation}\label{2}
\frac{da}{dt}=-\frac{\kappa}{2} a+\varepsilon
a^{\dagger}+gb+F_{c}(t),
\end{equation}
\begin{equation}\label{3}
\frac{db}{dt}=-\frac{\gamma }{2}b-ga+F_{e}(t),
\end{equation}
where $F_{c}$ and $F_{e}$ are the Langevin noise operators for the
cavity and exciton modes, respectively. Both noise operators have
zero mean, i.e., $\langle F_{c}\rangle=\langle F_{e}\rangle=0$. For
a cavity mode damped by a vacuum reservoir, the noise operator
satisfy the following correlations:
\begin{equation}\label{4}
\left\langle F_{c}(t)F_{c}^{\dagger}(t^{\prime })\right\rangle
=\kappa \delta (t-t^{\prime }),
\end{equation}
\begin{equation}\label{5}
\left\langle F_{c}^{\dagger}(t)F_{c}(t^{\prime })\right\rangle
=\left\langle F_{c}(t)F_{c}(t^{\prime })\right\rangle =\left\langle
F_{c}^{\dagger}(t)F_{c}^{\dagger}(t^{\prime })\right\rangle=0.
\end{equation}
The exciton noise operators satisfy the following correlations:
\begin{equation}\label{7}
\left\langle F_{e}(t)F_{e}^{\dagger}(t^{\prime })\right\rangle
=\gamma \delta (t-t^{\prime }),
\end{equation}
\begin{equation}\label{8}
\left\langle F_{e}^{\dagger}(t)F_{e}(t^{\prime })\right\rangle
=\left\langle F_{e}(t)F_{e}(t^{\prime })\right\rangle =\left\langle
F_{e}^{\dagger}(t)F_{e}^{\dagger}(t^{\prime })\right\rangle=0.
\end{equation}

\section{Photon statistics}
In this section we analyze the photon statistics of he fluorescent
light by calculating intensity, intensity spectrum and second order
correlation function in the strong coupling regime. The solution of
Eqs. \eqref{2} and \eqref{3} is rigorously derived in the Appendix.
In the paper paper is devoted to the dynamics of the system in the
strong coupling regime. To this end, imposing the strong coupling
limit ($g\gg\kappa,\gamma$), which amounts to keeping only the
leading terms in $g$, one obtains from Eqs. \eqref{A13a} and
\eqref{A17a} that $\Delta=\Lambda=4ig$. As a result, the solution
given by Eqs. \eqref{A18} and \eqref{A19} reduce to
\begin{align}\label{A18eq}
    a(t)&=
    \lambda_{1}^{(+)}(t)a(0)+\lambda_{2}^{(+)}(t)a^{\dagger}(0)+\lambda_{3}(t)b(0)+\lambda_{4}(t)b^{\dagger}(0)\notag\\
    &+\int_{0}^{t}dt^{\prime}~~\big[\lambda_{1}^{(+)}(t-t^{\prime})F_{c}(t^{\prime})+\lambda_{2}^{(+)}(t-t^{\prime})F_{c}^{\dagger}(t^{\prime})\big]\notag\\
    &+\int_{0}^{t}dt^{\prime}
    ~~\big[\lambda_{3}(t-t^{\prime})F_{e}(t^{\prime})+\lambda_{4}(t-t^{\prime})F_{e}^{\dagger}(t^{\prime})\big],
\end{align}
\begin{align}\label{A19eq}
    b(t)&=
    \lambda_{1}^{(-)}(t)b(0)+\lambda_{2}^{(-)}(t)b^{\dagger}(0)-\lambda_{3}(t)a(0)-\lambda_{4}(t)a^{\dagger}(0)\notag\\
    &-\int_{0}^{t}dt^{\prime}~~\big[\lambda_{3}(t-t^{\prime})F_{c}(t^{\prime})+\lambda_{4}(t-t^{\prime})F_{c}^{\dagger}(t^{\prime})\big]\notag\\
    &+\int_{0}^{t}dt^{\prime}
    ~~\big[\lambda_{1}^{(-)}(t-t^{\prime})F_{e}(t^{\prime})+\lambda_{2}^{(-)}(t-t^{\prime})F_{e}^{\dagger}(t^{\prime})\big],
\end{align}
where
\begin{align}\label{A20eq}
    \lambda_{1}^{(\pm)}(t)&=\Big[\Big(\cos (gt)
    \pm \frac{\gamma-\kappa}{4g}\sin(gt)\Big)\cosh (\varepsilon t/2)\notag\\
    &\pm\frac{\varepsilon}{2g}\sin (gt)\sinh(\varepsilon t/2)\Big]
    e^{-(\kappa+\gamma)t/4},
\end{align}
\begin{align}\label{A21eq}
    \lambda_{2}^{(\pm)}(t)&=\Big[\Big(\cos (gt)
    \pm \frac{\gamma-\kappa}{4g}\sin(gt)\Big)\sinh (\varepsilon t/2)\notag\\
    &\pm\frac{\varepsilon}{2g}\sin (gt)\cosh(\varepsilon
    t/2)\Big]e^{-(\kappa+\gamma)t/4},
\end{align}
\begin{equation}\label{A22eq}
   \lambda_{3}(t)=\sin (gt) \cosh (\varepsilon t/2)
   e^{-(\kappa+\gamma)t/4},
\end{equation}
\begin{equation}\label{A25eq}
   \lambda_{4}(t)=\sin (gt) \sinh (\varepsilon t/2)
   e^{-(\kappa+\gamma)t/4}.
\end{equation}
All quantities of interest which describe the dynamics of the system
can fully be analyzed using these solutions.

\subsection{Intensity of fluorescent light}
The dynamical behavior of the intensity of light emitted by a single
quantum well in GaAs microcavity has been measured experimentally
\cite{Jacobson95}. We here seek to study the dynamical behavior of
the light emitted by a single quantum well interacting with squeezed
light. The intensity of the fluorescent light is proportional to the
mean number of excitons in the system. Using Eq. \eqref{A19eq} and
the properties of the noise forces, we readily obtain
\begin{align}\label{10}
    \langle b^{\dagger}b\rangle &=
    \frac{2\varepsilon^2}{(\kappa+\gamma)^2-4\varepsilon^2}\notag+
    \frac{1}{2}\bigg[\big(1 + \bar n_{e} + \bar n_{e} \cos(2g t)\notag\\
    &+\frac{\kappa-\gamma}{4g}(1+2\bar n_{e})\sin(2gt)\big)\cosh (\varepsilon t) \notag\\&
    -\frac{1}{4g}[\kappa-\gamma+2\varepsilon(1 + 2 \bar n_{e})\sin(gt)]\sinh(\varepsilon t)\notag\\
    &-(\kappa+\gamma)\frac{2\varepsilon \sinh (\varepsilon t)+(\kappa+\gamma)\cosh(\varepsilon
    t)}{(\kappa+\gamma)^2-4\varepsilon^2}\notag\\
   &+\frac{\gamma-\kappa }{2g} \text{sinh}^2 (\varepsilon t/2)\text{sin}(2 g t)\bigg]e^{-(\kappa+\gamma)t/2},
\end{align}
where $\bar n_{e}$ is the mean exciton number in the cavity at
initial time. We assumed that the cavity mode is initially in vacuum
state. It is easy to see that in the steady state the mean exciton
number reduces to
\begin{align}\label{11}
    \langle b^{\dagger}(t)b(t)\rangle_{ss} =
    \frac{2\varepsilon^2}{(\kappa+\gamma)^2-4\varepsilon^2},
\end{align}
which is a contribution to intensity of the fluorescent light due to
the optical parametric oscillator.
\begin{figure}[h]
\includegraphics{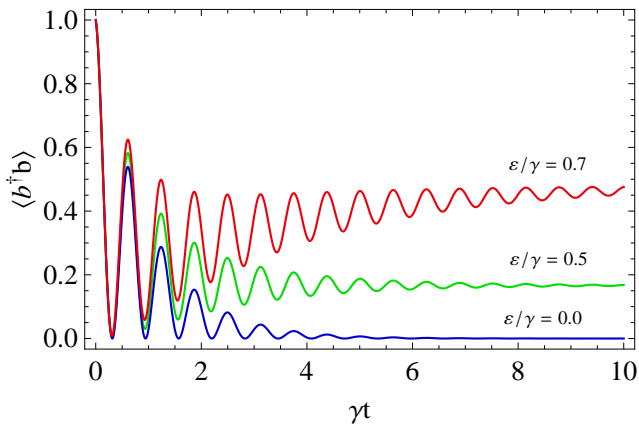}
\caption{Plots of the fluorescent intensity [Eq. \eqref{10}] vs
scaled time $\gamma t$ for $\gamma=\kappa$, $g/\gamma=5$, $\bar
n_{e}=1$ and for different values of $\varepsilon/\gamma$.}
\label{fig2}
\end{figure}
\begin{figure}[h]
\includegraphics{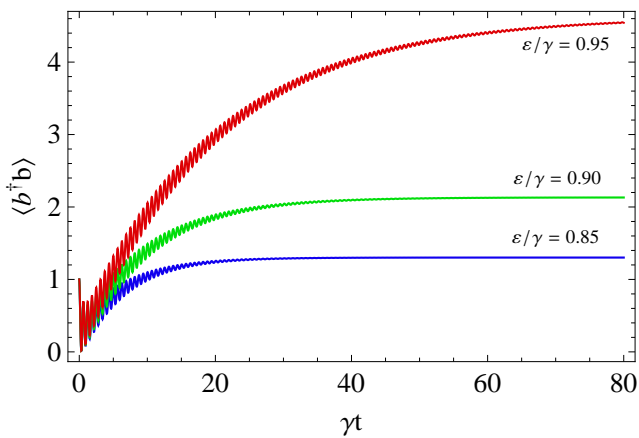}
\caption{Plots of the fluorescent intensity [Eq. \eqref{10}] near
threshold vs scaled time $\gamma t$ for $\kappa/\gamma=1$,
$g/\gamma=5$, $\bar n_{e}=1$ and for different values of
$\varepsilon/\gamma$.} \label{fig3}
\end{figure}

In Fig. 2, we  plot the intensity as a function of scaled time
$\gamma t$ for different values of the scaled pump field amplitude
$\varepsilon/\gamma$. In this figure we have assumed that the cavity
in initially prepared in such a way that it contains one
exciton($\bar n_{e}=1$) but no photon. For simplicity we have taken
the cavity and exciton decay rate to be the same, i.e.,
$\kappa=\gamma$. This figure shows the effect of the parametric
oscillator on the intensity fluorescent light. It is not hard to see
that the intensity oscillates with frequency equal to the coupling
constant $g$, which is a signature of exchange of energy between the
cavity and exciton modes. Moreover, the amplitude of the
oscillations depends on the amplitude of the pump field,
$\varepsilon$, which represents the optical parametric oscillator in
our system. The stronger the pump field and the higher the amplitude
of oscillation and the longer it takes to reach the steady state
value of the intensity.

It worth emphasizing that since optical parametric oscillator is
operating below threshold, the parameter $\varepsilon$ is
constrained by the inequality $\kappa+\gamma> 2\varepsilon$. We thus
interpret $\kappa+\gamma= 2\varepsilon$ as threshold condition for
the parametric process. In the vicinity of the threshold the mean
exciton number increases rapidly and exceeds unity as illustrated in
Fig. \ref{fig3}. This shows that even though there is one exciton in
the cavity initially, there is a finite probability for the squeezed
light in the cavity to excite two or more excitons in the quantum
well. This has an interesting effect on the photon statistics of the
fluorescent light as discussed in Section C.

\subsection{Intensity spectrum}
We next proceed to calculate the power spectrum of the fluorescent
light. The power spectrum of the fluorescent light can be expressed
in terms of the bosonic operator as
\begin{equation}\label{12}
    S(\omega)=\frac{1}{\pi}\text{Re}\int_{0}^{\infty}d\tau~e^{i\omega \tau}
    \frac{\langle b^{\dagger}(t)b(t+\tau)\rangle_{ss}}{\langle
    b^{\dagger}(t)b(t)\rangle_{ss}}.
\end{equation}
In the strong coupling regime the correlation function that appears
in the integrand of the power spectrum in the steady state has the
form
\begin{align}\label{14}
    \frac{\langle b^{\dagger}(t)b(t+\tau)\rangle_{ss}}{\langle
    b^{\dagger}(t)b(t)\rangle_{ss}}&=\Big[\frac{\gamma((\kappa+\gamma)^2-4\varepsilon^2)}{4g(\kappa+\gamma)\varepsilon}
    \sin(g\tau)\sinh(\varepsilon \tau/2)\notag\\
   &+\frac{\cos(gt)}{2\varepsilon}\big(2\varepsilon\cosh(\varepsilon \tau/2)\notag\\
   &+(\kappa+\gamma)\sinh(\varepsilon \tau/2)\big)\Big]
 e^{-(\kappa+\gamma)\tau/4}.
\end{align}
Substituting this result in Eq. \eqref{12} and keeping the leading
order in $g$, we obtain the power spectrum of the fluorescent light
to be
\begin{align}\label{15}
   S(\omega)&=\frac{\gamma_{+}\gamma_{-}}{2\pi\varepsilon g(\kappa+\gamma)}
  \Big[\frac{g\kappa+3g\gamma-2\gamma\omega}{\gamma_{-}^2+(g-\omega)^2}-\frac{g\kappa+3g\gamma-2\gamma\omega}{\gamma_{+}^2+(g-\omega)^2}\notag\\
  &+\frac{g\kappa+3g\gamma+2\gamma\omega}{\gamma_{-}^2+(g+\omega)^2}-\frac{g\kappa+3g\gamma+2\gamma\omega}{\gamma_{+}^2+(g+\omega)^2}\Big],
\end{align}
where $\gamma_{\pm}=(\gamma+\kappa\pm 2\varepsilon)/4$ are the half
widths of the Lorentzians centered at $\omega=\pm g$. We immediately
see that the width of the power spectrum depends on the amplitude of
the pump field.

\begin{figure}[h]
\includegraphics{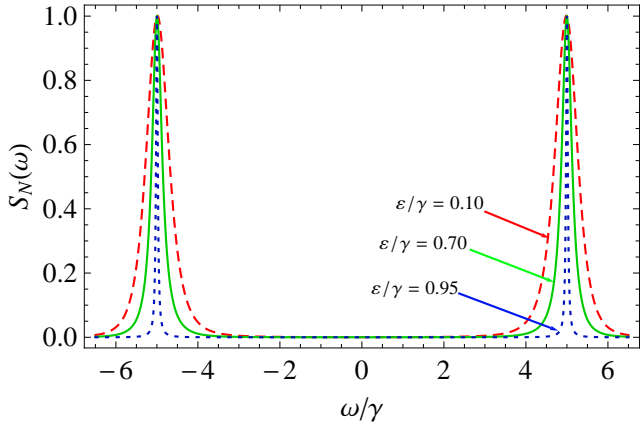}
\caption{Plots the normalized intensity spectrum of the fluorescent
light [$S_{N}(\omega)=S(\omega)/S(g)$] vs scaled frequency
$\omega/\gamma $ for $\kappa/\gamma=1$, $g/\gamma=5$, $\bar
n_{e}=1$, and for different values of $\varepsilon/\gamma$.}
\label{fig4}
\end{figure}
We observe that the maximum of the power spectrum occurs when the
frequency equal to the coupling constant ($g$). In order of explore
the effect of the squeezed light on the width of the spectrum it is
convenient to plot the the power spectrum normalized by its maximum
value, i.e., $S_{N}(\omega)=S(\omega)/S(g)$. In Fig.  \ref{fig4}, we
plot the normalized spectrum as a function of $\omega/\gamma$ for
different values of the pump amplitude ($\varepsilon$). As clearly
indicated in the figure, the the higher the amplitude of the pump
field (the degree of squeezing), the narrower the width has become.
It is also worth noting that the narrowing of the width is more
pronounced close to the threshold, i.e., when the squeezing
approaches to its maximum value. This is in contrary to the result
obtained when the quantum well is coupled to a squeezed vacuum
reservoir, where the spectrum is independent of the squeeze
parameter \cite{Erenso03}.

We further note that the spectrum has two peaks symmetrically
located at $\pm g$. This is the result of the strong coupling
approximation ($g\gg \kappa,\gamma$). Both peaks have the same width
which depends on the exciton and cavity modes decay rates and the
amplitude of the pump field.

\subsection{Autocorrelation function}
\begin{figure}[h]
\includegraphics{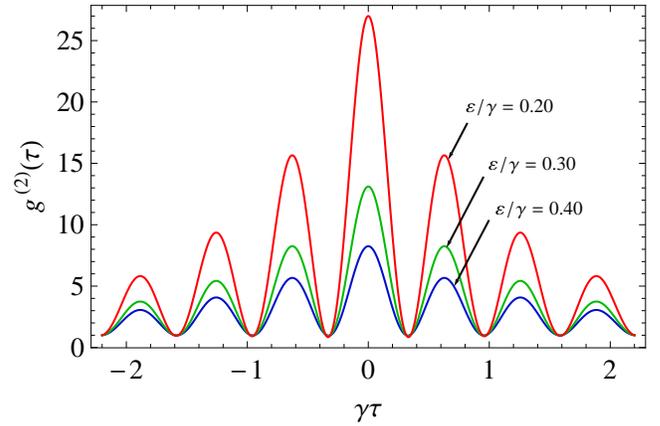}
\caption{Autocorrelation function versus normalized time $\gamma
\tau$ for $g/\gamma=5$, $\kappa/\gamma=1$, $\bar n_{e}=1$, and for
different value of pump amplitude $\varepsilon/\gamma$.}\label{g2}
\end{figure}

\ref{g2}
We now turn our attention to the calculation of autocorrelation
function, which is proportional to the probability of detecting one
photon at $t+\tau$ given that another photon was detected at earlier
time t. Quantum mechanically autocorrelation is defined by
\begin{equation}\label{G1}
g^{(2)}\left( \tau \right) =\frac{\left\langle b^{\dagger}
(t)b^{\dagger} (t+\tau )b(t+\tau )b(t)\right\rangle }{\left\langle
b^{\dagger} (t)b(t)\right\rangle ^{2}}.
\end{equation}
Using the Gaussian properties of the noise forces \cite{Walls94},
the autocorrelation function in the steady state can be put in a
simpler form
\begin{equation}\label{G2}
    g^{(2)}(\tau)=1+
    \frac{|\langle b^{\dagger }(t)b^{\dagger}(t+\tau)\rangle_{ss}|^2}{\langle b^{\dagger}(t)b(t)\rangle^2_{ss}}
    + \frac{|\langle b^{\dagger }(t)b(t+\tau)\rangle_{ss}|^2}{\langle
    b^{\dagger}(t)b(t)\rangle^2_{ss}}.
\end{equation}
In order to find a closed form analytical expression for the
autocorrelation function, one has to determine the two time
correlation functions that appear in Eq. \eqref{G1}. This can be
done using the solution \eqref{A19eq} along with the correlation
properties of the noise forces. After algebraic manipulations, we
obtain the final expression of the autocorrelation function to be
\begin{align}\label{G3}
g^{(2)}(\tau)=1+e^{-\frac{1}{2}(\kappa+\gamma)\tau}\cos(g\tau)\left[\mu_{1}\sin(g\tau)+\mu_{2}\cos(g\tau)\right],
\end{align}
where
\begin{align*}
\mu_{1}=\frac{\gamma((\kappa+\gamma)^2-4\varepsilon^2)}{4g(\kappa+\gamma)\varepsilon^2}\left[(\kappa+\gamma)\cosh(\varepsilon
\tau)+2\varepsilon\sinh(\varepsilon \tau) \right],
\end{align*}
\begin{equation*}
\mu_{2}=\frac{((\kappa+\gamma)^2+4\varepsilon^2)\cosh(\varepsilon
\tau)+4(\kappa+\gamma)\varepsilon \sinh(\varepsilon
\tau)}{4\varepsilon^2}.
\end{equation*}
Expression \eqref{G3} is valid only in the strong coupling regime
($g\gg\kappa,\gamma$).

The behavior of $g^{(2)}(\tau)$ as a function of the pump amplitude
($\varepsilon$) and for constant $g$ is illustrated in Fig.
\ref{g2}. This figure shows that the correlation function oscillates
at frequency equals to $g$. The amplitude of this oscillations
decreases fast when we increase the value of $\varepsilon$. The
autocorrelation function at $\tau=0$ has the form
$g^{(2)}(0)=2+(\kappa+\gamma)^2/4\varepsilon^2>1$ indicating the
phenomenon of photon bunching. Here the underlying physics can be
explained in terms of the mean exciton number (see Fig. \ref{fig3}).
In that figure we have showed that, even though we start of one
exciton initially, there is finite probability of exciting two or
more excitons in the quantum well by the squeezed light. This allows
the possibility of emission of two photon at a time which leads to
of phenomenon of bunching in the fluorescent light.

\section{Quadrature squeezing}
The squeezing properties of the fluorescent light can be analyzed by
calculating the variances of the quadrature operators. The variances
of the quadrature operators for the fluorescent light are given by
\begin{equation}\label{Q1}
    \Delta b_{1}^2=1+2\langle b^{\dagger}b\rangle+\langle
    b^{2}\rangle+\langle b^{\dagger 2}\rangle,
\end{equation}
\begin{equation}\label{Q2}
    \Delta b_{2}^2=1+2\langle b^{\dagger}b\rangle-(\langle
    b^{2}\rangle+\langle b^{\dagger 2}\rangle),
\end{equation}
where $b_{1}=b^{\dagger}+b$ and $b_{2}=i(b^{\dagger}-b)$. These
quadrature operators satisfy the commutation relation
$[b_{1},b_{2}]=2i$. On the basis of these definitions the
fluorescent light is said to be in a squeezed state if either
$\Delta b_{1}^2<1$ or $\Delta b_{2}^{2}<1$. In deriving $\eqref{Q1}$
and $\eqref{Q2}$ we have used $\langle b(t)\rangle=0$, which can
easily be verified using $\eqref{A19eq}$. Applying Eq. \eqref{A19eq}
and the properties of the noise operators the variances turn out to
be
\begin{align}\label{Q3}
    \Delta b_{1}^2=1&+\frac{2 \varepsilon }{k+\gamma -2 \varepsilon
    }+e^{-(k+\gamma -2\varepsilon)t/2}A_{-}(t)\notag\\
    &+e^{-(k+\gamma -\varepsilon)t/2} B_{-}(t),
\end{align}
\begin{align}\label{Q4}
    \Delta b_{2}^2=1&-\frac{2 \varepsilon }{k+\gamma +2 \varepsilon
    }+e^{-(k+\gamma +2\varepsilon)t/2}A_{+}(t)\notag\\
    &+e^{- (k+\gamma +\varepsilon)t/2} B_{+}(t),
\end{align}
in which
\begin{align*}
    A_{\pm}(t)&=1+\bar n_{e}+\bar n_{e}\cos(2gt)\notag\\
    &\Big[\frac{\gamma-\kappa}{4g}e^{\pm \varepsilon t}+\frac{\kappa-\gamma\pm 2\varepsilon}{4g}(1+2\bar
    n_{e})\Big]\sin(2gt),
\end{align*}
\begin{align*}
    B_{\pm}(t)=-\frac{(\kappa+\gamma)e^{\pm \varepsilon
    t/2}}{\kappa+\gamma\pm 2\varepsilon}\pm \frac{\kappa-\gamma}{2g}\sinh(\varepsilon
    t/2)\sin(2gt).
\end{align*}
It is straightforward to see that the variances reduce in the steady
state to
\begin{align}\label{Q5}
\Delta b_{1}^2&=1+\frac{2 \varepsilon }{k+\gamma -2 \varepsilon},
\end{align}
\begin{align}\label{Q6}
\Delta b_{2}^2&=1-\frac{2 \varepsilon }{k+\gamma +2 \varepsilon}.
\end{align}
Expressions \eqref{Q5} and \eqref{Q6} represent the quadrature
variance of a parametric oscillator operating below threshold.  At
threshold $\kappa+\gamma=2\varepsilon$, the squeezing becomes $50\%$
which is the maximum squeezing that can be obtained from
subthreshold parametric oscillator \cite{Milburn83}. It is then not
difficult to see that the squeezing occurs in the $b_{2}$
quadrature.

\begin{figure}[h]
\includegraphics{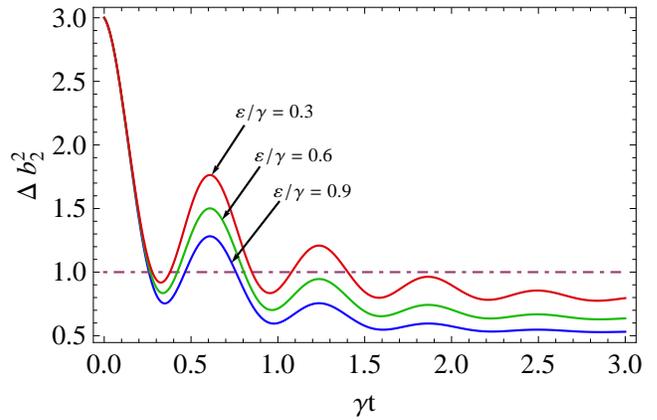}
\caption{Plots of the quadrature variance [Eq. \eqref{Q4}] vs scaled
time $\gamma t$ for $g/\gamma=5$, $\kappa=\gamma$, $\bar n_{e}=1$
and for the different values of the pump field amplitude
$\varepsilon/\gamma$.}\label{qv}
\end{figure}

In Fig. \ref{qv}, the time evolution of the variance of the $b_{2}$
quadrature, \eqref{Q4}, is plotted versus scaled time $\gamma t$.
The variance in this quadrature oscillates with frequency equal to
twice the Rabi frequency. The amplitude of oscillation damps out at
longer time and eventually become flat at steady state. Moreover, it
is interesting to note that the fluorescent light is not squeezed at
initial moment however, it starts to exhibit transient squeezing
before it becomes unsqueezed again. The more the exciton interacts
with the squeezed light, the stronger the squeezing becomes. As a
result of this we observe squeezed fluorescent light in longer
periods which ultimately approaches to the $50\%$ maximum squeezing
limit observed in parametric oscillator. The reduction of
fluctuations noted in the fluorescent light is due to the
interaction between the long-lived squeezed photons in the cavity
and excitons in the quantum well. As can be seen from Fig. \ref{qv},
the degree of squeezing of the fluorescent light depends on the
amplitude of the pump field.

\section{Conclusion}
The quantum statistical properties of the fluorescent light emitted
by exciton in a quantum well interacting with squeezed light is
presented. Analytical solutions for the pertinent quantum Langevin
equations are rigorously derived. These solutions, in the strong
coupling limit in which the exciton-cavity mode coupling is much
greater than the cavity as well as exciton spontaneous decay rates
$(g\gg \kappa,\gamma)$ are used to study the dynamical behavior of
the generated light. We find that the squeezed light enhances the
mean photon number and narrows the width of the intensity spectrum
of the fluorescent light. Further, the fluorescent light shows
normal-mode splitting, which is a signature of strong coupling. We
note that unlike atomic cavity QED where the fluorescent light
exhibits antibunching, the fluorescent light in the present system
rather exhibits bunching. The manifestation of bunching is
attributed to the possibility of exciting two or more excitons in
the quantum well which in turn leads a finite probability of
emission of two photons simultaneously.

\begin{acknowledgements}
One of us (E.A.S) gratefully acknowledge financial support from the
Robert A. Welch and the Heep Foundations.
\end{acknowledgements}

\appendix
\section{Solution for the quantum Langevin equations}
In this appendix we derive the solution of the following quantum
Lagevin equations:
\begin{equation}\label{A1}
\frac{da}{dt}=-\frac{\kappa}{2} a+\varepsilon
a^{\dagger}+gb+F_{c}(t),
\end{equation}
\begin{equation}\label{A2}
\frac{db}{dt}=-\frac{\gamma }{2}b-ga+F_{e}(t).
\end{equation}
In order to solve these equations it is more convenient to introduce
new variable defined by
\begin{equation}\label{A3}
    a_{\pm}=a^{\dagger}\pm a ~~~~~~~~~b_{\pm}=b^{\dagger}\pm b.
\end{equation}
With the help of Eqs. \eqref{A1} and \eqref{A2} and their complex
adjoint we obtain
\begin{eqnarray}
  \frac{d}{dt}a_{+} &=& -\frac{1}{2}(\kappa-2\varepsilon)a_{+}+gb_{+}+F_{+} \label{A4}\\
  \frac{d}{dt}b_{+} &=&
  -\frac{\gamma}{2}b_{+}-ga_{+}+G_{+}\label{A5}
\end{eqnarray}
\begin{eqnarray}
  \frac{d}{dt}a_{-} &=& -\frac{1}{2}(\kappa+2\varepsilon)a_{-}+gb_{-}+F_{-}\label{A6} \\
  \frac{d}{dt}b_{-} &=&
  -\frac{\gamma}{2}b_{-}-ga_{-}+G_{-},\label{A7}
\end{eqnarray}
where $F_{\pm}=F_{c}^{\dagger}\pm F_{c}$ and
$G_{\pm}=F_{e}^{\dagger}\pm F_{e}$. Note that Eqs. \eqref{A4} and
\eqref{A5} are decoupled from \eqref{A6} and \eqref{A7}. These
coupled equations can be solved using the method of Laplace
transform.

The Laplace transform of Eqs. \eqref{A4} and \eqref{A5} gives
\begin{align}\label{A8}
    A(s)&=\frac{4g}{\chi}G(s)+\frac{2(2s+\gamma)}{\chi}F(s)\notag\\
    &+\frac{1}{\chi}\big[4gb_{+}(0)+2(2s+\gamma)a_{+}(0)\big]
\end{align}
\begin{align}\label{A9}
    B(s)&=\frac{2}{\chi}(\kappa+2s-2\varepsilon)G(s)-\frac{4g}{\chi}F(s)\notag\\
    &+\frac{1}{\chi}\big[-4ga_{+}(0)+2(\kappa+2s-2\varepsilon)b_{+}(0)\big],
\end{align}
where $\chi=4g^2+(2s+\gamma)(\kappa+2s-2\varepsilon)$ and
$A(s)=\mathcal{L}(a_{+}), B(s)=\mathcal{L}(b_{+}),
G(s)=\mathcal{L}(G_{+}) $ and $F(s)=\mathcal{L}(F_{+})$ with
$\mathcal{L}$ denoting Laplace transform. The inverse Laplace
transform of Eqs. \eqref{A8} and \eqref{A9} yields
\begin{align}\label{A10}
    a_{+}(t)&=a_{+}(0)f_{+}(t)
    +b_{+}(0)f_{2}(t)+\int_{0}^{t}f_{+}(t-t^{\prime})F_{+}(t^{\prime})dt^{\prime}\notag\\
    &+\int_{0}^{t}f_{2}(t-t^{\prime}) G_{+}(t^{\prime})dt^{\prime}
\end{align}
\begin{align}\label{A11}
    b_{+}(t)&=b_{+}(0)f_{-}(t)-
    a_{+}(0)f_{2}(t)+\int_{0}^{t}f_{-}(t-t^{\prime})G_{+}(t^{\prime})dt^{\prime}\notag\\
    &-\int_{0}^{t}f_{2}(t-t^{\prime}) F_{+}(t^{\prime})dt^{\prime},
\end{align}
where
\begin{equation}\label{A12}
    f_{\pm}(t)=\left[\cosh
    (\Delta t/4)\pm \frac{\gamma-\kappa+ 2\varepsilon}{\Delta}\sinh(\Delta t/4)\right]e^{-\gamma_{-}t}
\end{equation}
\begin{equation}\label{A13}
    f_{2}(t)=\frac{4g}{\Delta}\sinh (\Delta t/4) e^{-\gamma_{-}t}.
\end{equation}
\begin{equation}\label{A13a}
    \Delta=\sqrt{-16g^2+(\gamma-\kappa+ 2\varepsilon)^2},~~
    \gamma_{-}=\frac{1}{4}(\kappa+\gamma-2\varepsilon)
\end{equation}

Note that the solution of the coupled equations \eqref{A6} and
\eqref{A7} can easily be obtained by replacing $\varepsilon$ by
$-\varepsilon$, $F_{+}$ by $F_{-}$, and $G_{+}$ by $G_{-}$ in the
solution of Eqs. \eqref{A4} and \eqref{A5}. We thus have
\begin{align}\label{A14}
    a_{-}(t)&=a_{-}(0)h_{+}(t)
    +b_{-}(0)h_{2}(t)+\int_{0}^{t}h_{+}(t-t^{\prime})F_{-}(t^{\prime})dt^{\prime}
    \notag\\&+\int_{0}^{t}h_{2}(t-t^{\prime}) G_{-}(t^{\prime})dt^{\prime}
\end{align}
\begin{align}\label{A15}
    b_{-}(t)&=b_{-}(0)h_{-}(t)-
    a_{-}(0)h_{2}(t)+\int_{0}^{t}h_{-}(t-t^{\prime})G_{-}(t^{\prime})dt^{\prime}\notag\\
    &-\int_{0}^{t}h_{2}(t-t^{\prime}) F_{-}(t^{\prime})dt^{\prime},
\end{align}
where
\begin{equation}\label{A16}
    h_{\pm}(t)=\left[\cosh
    (\Lambda t/4)\pm \frac{\gamma-\kappa- 2\varepsilon}{\Lambda}\sinh (\Lambda t/4)\right]e^{-\gamma_{+}t}
\end{equation}
\begin{equation}\label{A17}
    h_{2}(t)=\frac{4g}{\Lambda}\sinh (\Lambda t/4) e^{-\gamma_{+}t}.
\end{equation}
\begin{equation}\label{A17a}
    \Lambda=\sqrt{-16g^2+(\gamma-\kappa- 2\varepsilon)^2},~~
    \gamma_{+}=\frac{1}{4}(\kappa+\gamma+2\varepsilon)
\end{equation}

Applying the inversion formula $a=(a_{+}-a_{-})/2$ and
$b=(b_{+}-b_{-})/2$ the solution for $a(t)$ and $b(t)$ turn out to
be
\begin{align}\label{A18}
    a(t)&=
    \eta_{1}^{(+)}(t)a(0)+\eta_{2}^{(+)}(t)a^{\dagger}(0)+\eta_{3}^{(+)}(t)b(0)+\eta_{3}^{(-)}(t)b^{\dagger}(0)\notag\\
    &+\int_{0}^{t}dt^{\prime}~~\big[\eta_{1}^{(+)}(t-t^{\prime})F_{c}(t^{\prime})+\eta_{2}^{(+)}(t-t^{\prime})F_{c}^{\dagger}(t^{\prime})\big]\notag\\
    &+\int_{0}^{t}dt^{\prime}
    ~~\big[\eta_{3}^{(+)}(t-t^{\prime})F_{e}(t^{\prime})+\eta_{3}^{(-)}(t-t^{\prime})F_{e}^{\dagger}(t^{\prime})\big],
\end{align}
\begin{align}\label{A19}
    b(t)&=
    \eta_{1}^{(-)}(t)b(0)+\eta_{2}^{(-)}(t)b^{\dagger}(0)-\eta_{3}^{(+)}(t)a(0)-\eta_{3}^{(-)}(t)a^{\dagger}(0)\notag\\
    &-\int_{0}^{t}dt^{\prime}~~\big[\eta_{3}^{(+)}(t-t^{\prime})F_{c}(t^{\prime})+\eta_{3}^{(-)}(t-t^{\prime})F_{c}^{\dagger}(t^{\prime})\big]\notag\\
    &+\int_{0}^{t}dt^{\prime}
    ~~\big[\eta_{1}^{(-)}(t-t^{\prime})F_{e}(t^{\prime})+\eta_{2}^{(-)}(t-t^{\prime})F_{e}^{\dagger}(t^{\prime})\big],
\end{align}
where
\begin{align}\label{A20}
    \eta_{1}^{(\pm)}(t)&=\frac{1}{2}\Big(\cosh (\Delta t/4)
    \pm \frac{\gamma-\kappa+2\varepsilon}{\Delta}\sinh(\Delta t/4)\Big)e^{-\gamma_{-}t}\notag\\
    &+\frac{1}{2}\Big(\cosh (\Lambda t/4)
    \pm \frac{\gamma-\kappa-2\varepsilon}{\Lambda}\sinh(\Lambda t/4)\Big)e^{-\gamma_{+}t}
\end{align}
\begin{align}\label{A21}
    \eta_{2}^{(\pm)}(t)&=\frac{1}{2}\Big(\cosh (\Delta t/4)
    \pm \frac{\gamma-\kappa+2\varepsilon}{\Delta}\sinh(\Delta t/4)\Big)e^{-\gamma_{-}t}\notag\\
    &-\frac{1}{2}\Big(\cosh (\Lambda t/4)
    \pm \frac{\gamma-\kappa-2\varepsilon}{\Lambda}\sinh(\Lambda t/4)\Big)e^{-\gamma_{+}t}
\end{align}
\begin{equation}\label{A22}
   \eta_{3}^{(\pm)}(t)=\frac{2g}{\Delta}\sinh (\Delta
   t/4)e^{-\gamma_{-}t}\pm \frac{2g}{\Lambda}\sinh (\Lambda
   t/4)e^{-\gamma_{+}t}.
\end{equation}

\end{document}